\documentclass[aps,prl,twocolumn]{revtex4}
\usepackage[latin1]{inputenc}
\newcommand{\average}[1]{\left<{#1}\right>}

\begin{document}

\title{Comment on ``Failure of the work-hamiltonian connection for
free-energy calculations'' by José M. G. Vilar and J. Miguel Rubi}

\author{Luca Peliti}

\affiliation{Dipartimento di Scienze Fisiche,  Unità CNISM and
Sezione INFN, Università ``Federico II'', Complesso Monte S. Angelo,
80126 Napoli, Italy}

\date{July 29, 2008}

\maketitle

If the arguments put forward by Vilar and Rubi in their recent
work~\cite{VR1} were valid, quite a few accepted results in standard
statistical mechanics would have to be revised. Let us consider a
system with the hamiltonian $H(x,\lambda)=H_0(x)-Q(x,\lambda)$, in
which $\lambda$ is a parameter,  initially at equilibrium at the
inverse temperature $\beta$. Its free energy, according to the usual
understanding of statistical mechanics
(\cite[(484)]{Gibbs},\cite[(133.1--2)]{Tolman}) is given by
\begin{equation}\label{partition:eq}
    G_\lambda=-\frac{1}{\beta}\ln Z_\lambda,
\end{equation}
where $Z_\lambda=\int dx\; e^{-\beta H(x,\lambda)}$, and the
integral runs over all the microscopic states of the system. Thus,
if the parameter $\lambda$ changes from $\lambda_0$ to $\lambda_1$
and the system is in equilibrium both at the beginning and at the
end at the inverse temperature $\beta$, its free energy change
should be given by $\Delta
G=-\beta^{-1}\ln\left[Z_{\lambda_1}/Z_{\lambda_0}\right]$. According
to Vilar and Rubi~\cite{VR1}, this expression for $\Delta G$ ``is
not thermodynamically valid when changes of the Hamiltonian cannot
be associated with the work performed on the system''. If they are
right, since the expression for $\Delta G$ follows from
(\ref{partition:eq}) by subtraction, the connection
(\ref{partition:eq}) between the free energy and the partition
function, which is a cornerstone of the statistical mechanics
interpretation of thermodynamics, is not valid either. Let us point
out that the above expression of the free energy change is a direct
consequence of the thermodynamical relation $\Delta
G=\Delta(E-TS)=W^{\mathrm{th}}$, valid for reversible isothermal
transformations, and of the standard expression
(\cite[p.~42--44]{Gibbs},\cite[p.~527--535]{Tolman}) of the
thermodynamical work, $W^{\mathrm{th}}=\int_{\lambda_0}^{\lambda_1} d\lambda\;
\average{\partial H/\partial \lambda}_\lambda$, where 
$\average{A}_\lambda=\int dx
\,A(x)\,e^{-\beta H(x,\lambda)}/Z_\lambda$ is the canonical average with the hamiltonian $H(x,\lambda)$. (See in
particular~\cite[(121.8), p.~535; (124.1), p.~542]{Tolman}.) Note
moreover that, if an ergodic system undergoes an infinitely slow
parameter change, the time integral $W=\int dt\;\partial H(x(t),
\lambda(t))/\partial \lambda\,\dot\lambda(t)$ is equal to $W^{\mathrm{th}}$
in any realization of the process, independently of the size
of the system. Thus it is natural to define $W$ as the fluctuating work, which
is equal to $W^\mathrm{th}$ in an infinitely slow process.  This
quantity satisfies a number of important fluctuation relations,
in particular the Jarzynski equality (JE), $\average{e^{-\beta W}}=Z_{\lambda1}/Z_{\lambda_0}=e^{-\beta\,\Delta G}$, where the angular brackets denote the average
with respect to all realizations of the process.~\cite{Jarzynski1}
Vilar and Rubi contend that the
time-honored statistical mechanics expression of the thermodynamical
work reported above is incorrect. They maintain that $W$
is a recently introduced \emph{ad~hoc}
redefinition of work, which ``does not solve the physical inconsistencies,
such as the dependence of $\Delta G_Z$ on arbitrary parameters''.
(The fact that these ``physical inconsistencies'' are illusory has
been discussed elsewhere~\cite{Peliti}.) Vilar and Rubi prescribe
that one should consider, instead of $W$, the work performed on the
system during a manipulation, given by $W_0=\int dt \;\dot
x(t)\,\partial Q(x(t),\lambda(t))/\partial x$, and that $W_0$ does
not satisfy the Jarzynski equality, but, e.g., satisfies
$\average{e^{-\beta W_0}}=1$ for the case of a sudden change of the
Hamiltonian. This last identity is indeed correct, and is a special
case of an identity noticed long ago by Bochkov and
Kuzovlev~\cite{Bochkov}. However, it does not affect the JE, which
holds for $W$. Indeed, $W$ does not represent the work done on the
system, but rather the work done by the system on the external
bodies which produce the change of the hamiltonian (as emphasized
by Gibbs~\cite[p.~42]{Gibbs} and Tolman~\cite[p.~530, 2nd paragraph]{Tolman}). 
The two works are in general different, and it is $W$ that is related to
the thermodynamical work. The connection between the two works and
their fluctuation relations has been recently discussed in detail by
Jarzynski~\cite{Jarzynski2}. When the system evolves over a finite
time interval, we have in general
$W_0-W=Q(x(t),\lambda(t))-Q(x(0),\lambda(0))$.
It turns out that $W$ is more useful than $W_0$ for the
reconstruction of free-energy landscapes, because one has not yet
identified identities satisfied by $W_0$ which could be applied in
this context. In any case, renouncing the use of 
$W$ would entail giving
up the connection (\ref{partition:eq}) between the free energy and
the partition function, and
would therefore require an extensive rewriting of the basic
principles of statistical mechanics.

\end{document}